# HEURISTIC APPROACH OF AUTOMATED TEST DATA GENERATION FOR PROGRAMS HAVING ARRAY OF DIFFERENT DIMENSIONS AND LOOPS WITH VARIABLE NUMBER OF ITERATION

Hitesh Tahbildar[1] and Bichitra Kalita[2]

[1]Department of Computer Engineering and Application, Assam Engineering Institute, Guwahati, Assam 781003, India

tahbil@rediffmail.com

[2]Department of Computer Application, Assam Engineering College, Guwahati, Assam 781003, India

bichitra1_kalita@rediffmail.com

## ABSTRACT

*Normally, program execution spends most of the time on loops. Automated test data generation devotes special attention to loops for better coverage. Automated test data generation for programs having loops with variable number of iteration and variable length array is a challenging problem. It is so because the number of paths may increase exponentially with the increase of array size for some programming constructs, like merge sort. We propose a method that finds heuristic for different types of programming constructs with loops and arrays. Linear search, Bubble sort, merge sort, and matrix multiplication programs are included in an attempt to highlight the difference in execution between single loop, variable length array and nested loops with one and two dimensional arrays. We have used two parameters/heuristics to predict the minimum number of iterations required for generating automated test data. They are longest path level ($k_L$) and saturation level ($k_S$). The proceedings of our work includes the instrumentation of source code at the elementary level, followed by the application of the random inputs until all feasible paths or all paths having longest paths are collected. However, duplicate paths are avoided by using a filter. Our test data is the random numbers that cover each feasible path.*

## 1. INTRODUCTION

In software testing, loops are important spot for error detection. Execution of program spend large amount of time in loops. Without covering paths going through loops we can not get better code coverage. Most of the mistakes are made in loops of programs. Infinite loop creates lots of problem in detecting the errors. In fact, it is impossible to detect all kinds of infinite looping automatically [23]. Test data generation is more challenging if loops are nested. Automated test data is generated using symbolic value, actual value, and combining both. One of the main problems in test data generation is detection of infeasible path. Statistics reveals that many paths of a program can be infeasible[5]. The symbolic execution method suffers for infeasible path detection due to non availability of efficient constraint solver and path feasibility detector. The actual value execution method may spend lot of computation to detect the infeasible path. It is observed that combined approach [16] is better method for avoiding infeasible path. In a loop many paths are infeasible. It is seen that some combined method [16] does code instrumentation and constraint solving. It has been found that [16], the PathCrawler prototype tool is a more convenient method for automatic test data generation of programs having loops and array constructs. But the number of iteration applied to loop construct is satisfied with a minimum value like $k = 2$ where $k$ is number of iteration allowed for finding paths for test data generation. This is because the number of paths increases exponentially for programming construct like merge sort where we have two array of variable length and three sequential loops. The value of $k$ is determined by trial and error method. The authors proposed a heuristic called longest path criteria instead of $k = 2$ path criterion to improve automated test data generation for programs having arrays of variable length, and loops with variable number of iteration [12]. The methods ignore infeasible path problem as it considers only the feasible paths. Test data generation for programs with loops causes a combinatorial explosion in the number of execution paths for programming construct like merge sort [12]. But it is seen that the number of feasible path decreases in case of matrix multiplication program because of the restriction of equal number of rows and columns. The number of path is linearly increased with increase of array size for program with one loop statement like linear search. The number of path increases quadratically for bubble sort program with increase of array size. The basic idea of our work is taken from path crawler. [16]. The path crawler work includes instrumentation of the source code so as to recover the symbolic execution path each time that the program under test is executed. The code is first executed using inputs arbitrarily selected from the input domain. The resulting symbolic path is transformed into a path-predicate by projection of conditions into the input variables. They have achieved the next test using constraint logic programming to find new input values outside the domain of the path already covered. The instrumented code is then executed on this test and so on until all the feasible paths have been covered. In our work we avoid constraint solving part by generating random inputs to the instrumented code until all feasible paths are covered. That is we stop when no more new feasible path is created. This is saturation level($k_S$). For some programs with large array size, like merge sort it may not be possible to get saturation level, because the number of paths increases exponentially when array size is increased. In that case our coverage criteria will be $k_L$.( minimum number of iteration where longest path exist). Our experimental results confirm that after certain number of iteration we found all the feasible paths. Even if we increase the number of iteration no more paths can be generated. The randomly generated inputs may cover same path more that one time as we know there can be more than one input that traverse the same path. The paths generated by our random inputs are passed through a filter to find out the unique feasible paths. An important issue of our method is how long we will continue random number generation. We collected from our experiments a number of test cases, unique feasible path(UFP), new feasible path(NFP), longest length path(LLP), and execution time(Etime) for different number of iteration(k). Our experimental results shows that for program like linear search no more new path is generated if we take number of iteration greater than array size, for programs like matrix multiplication number of iteration is greater than $a3$ where $a$ is number of rows/column of a square matrix. Number of iteration for program like bubble sort can be taken as half of square of array size. For programs like merge sort we can not get $k_S$ level as the number of paths increases exponentially with the increase of array size. In this case, we take longest path level($k_L$)[12]. $k_L$ increases linearly with the increase of array size. Similarly we can have a heuristic table for other programming constructs those can be used for determining the number of iterations to be adopted for test data generation. Therefore our test data generation method provide us test data for programs with variable number of loops and array length in *less effort* and with *better coverage* in comparison to path crawler method.

The rest of the paper is organized as follows: The section 2 presents a survey of related works of path oriented test data generation. Section 3 describes our approach of test data generation. Section 4 illustrates our test data generation process with examples. Section 5 shows our experiment results and propose heuristics for different types of programming constructs. Section 6 discusses our experimental results. Finally in section 7 we conclude with some observation and future works for automatic test data generation of loop and array constructs.

## 2. Related Work

In [18] Prather described a new software testing strategy. The method uses previous test paths to guide in the subsequent path selection. The method ensures branch coverage and generates path dynamically making the best possible use of previously generated test cases. But the method does not give clear cut idea about infeasible paths. Flow analysis can identify infeasible paths. Gustafsson etc. [14, 13] presented method that uses static worst case Execution time (WCET) analysis to derive upper bounds of nested loops and automatic detection of infeasible path using abstract execution. But the claim of improvement of WCET is only for some programming constructs. Williams [3] proposed a novel method for the automatic test data generation of tests satisfying the *k* paths criteria. The method claims 100% coverage of feasible paths, but uses trial and error method to predict the value of *k*. The method requires code instrumentation and constraint solving. Williams present a Pathcrawler tool [16] for automatic generation of test cases satisfying user defined limit. Their method is a combination of both static and dynamic analysis in a way that avoids the disadvantages of both. The authors could not satisfy the all paths criterion. Their approach was limited to k-path (k being the number of user defined loop iterations). The value of *k* is restricted to two for avoiding exponential increase of paths. They have used a trial and error method for computing *k*. They failed to put forth an upper bound testing. Approach was constricted only to merge-sort program. Hence diversified observation of the statistical variation of data could not be observed and it proved very difficult to come to a generalized conclusion regarding their observations. In [12] a heuristic is developed based on experimental results to predict the value of *k* for which longest path is covered. The heuristic avoids trial and error method of predicting the minimum number of iteration *k*. But it is focused on longest path criteria for predicting *k*. All feasible paths are not covered in longest path criteria. There are many programming construct where saturation level ($k_S$) and longest path level ($k_L$) difference is minimal and that can be neglected for coverage analysis.

## 3. Our Approach

### 3.1. Model

In our model, we first instrument source code so as to print out the symbolic execution paths. Then we apply random inputs from a given domain to the instrumented object code for extracting all possible feasible paths. All paths are collected and then filtered to get unique feasible paths. Main issue in our model is after how many iteration all feasible paths are collected. The different phases of our approach are shown in figure 1. This approach is applicable to all sequential programs coded in an imperative language and the prototype has been implemented for C using function merge sort, linear search, bubble sort and matrix multiplication.

It starts with the instrumentation of the source code. The instrumentation stage is an automatic transformation of the source code to a form that can print out all the feasible paths when random inputs are supplied by random number generator within a defined domain. The random inputs will be supplied to the instrumented code until no more new paths are generated. The number of iteration required is denoted as $k_L$ and $k_S$. There can be duplicate paths. Because random data may traverse same path more than once. The paths generated are filtered to get the unique feasible paths by using shell script. The comparator is used to compare either $k_L$ or $k_S$ value depending on programming constructs. The test data generation process terminates when the difference of $k_L/k_S$ between current and previous iteration is equal to zero. Our test data is the random numbers that cover the unique feasible paths.

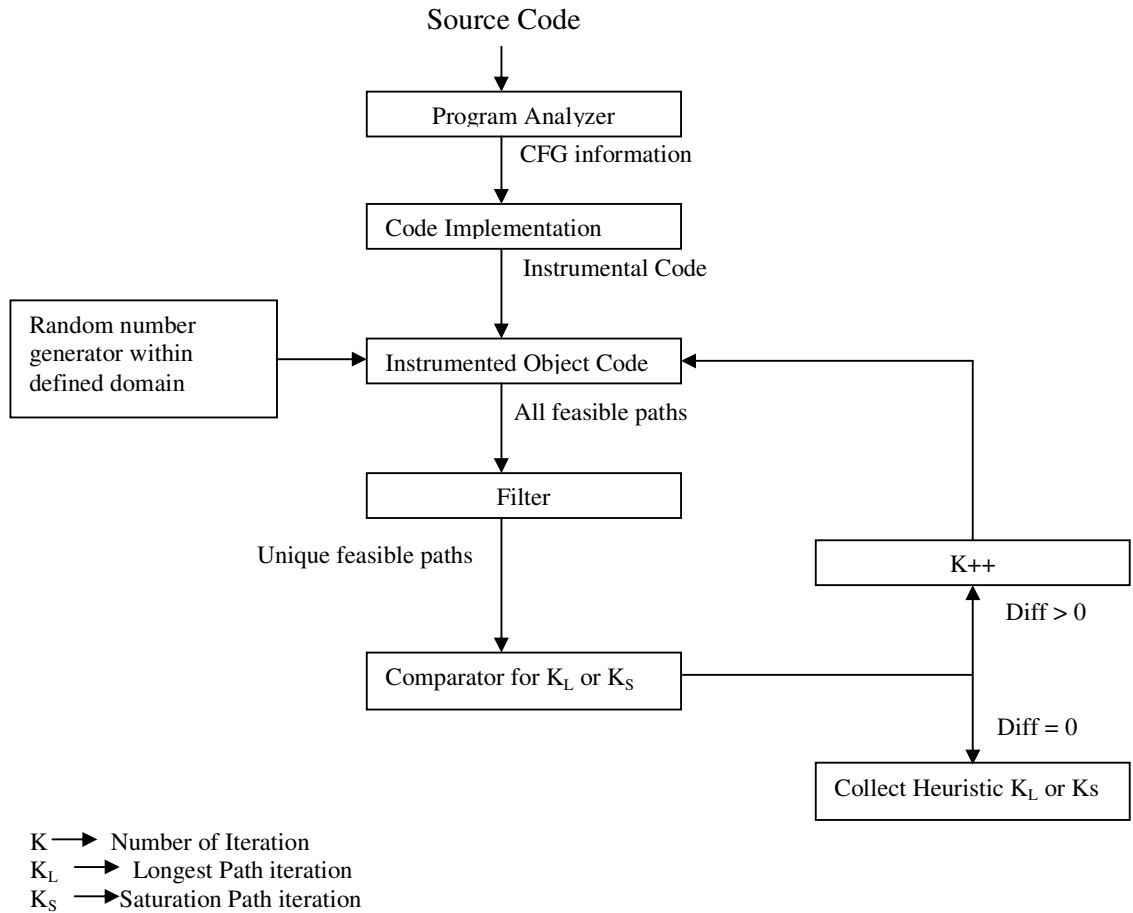

K → Number of Iteration
$K_L$ → Longest Path iteration
$K_S$ → Saturation Path iteration

Figure 1: Sequential overview of the process

### .3.2. Test Data

Our test data are random numbers that cover unique feasible paths. There may be many inputs that cover the same path. We take one input for each unique path which is our test data.

## 4. Examples

### 4.1. Matrix multiplication

Matrix multiplication takes as input two matrices *A*[*a*1*; a*2] and *B*[*a*3*; a*4] of dimensions (*a*1*; a*2) and (*a*3*; a*4) respectively, where *a*2 = *a*3 and produces the output in another matrix *C*[*a*1*; a*4]. Input: The input of the program is received from the user includes the following –

Maximum limit of dimensions for the two matrices a[ ][ ] and b[ ][ ]. Maximum range of domain from which elements of the matrices are randomly selected. Maximum number of iterations to be allowed during feasible path generation. Source Code of the function Matrix multiplication is given in Appendix I. The control flow of the graph is shown in figure 2. Test generated for Matrix: Considering two matrices a and b of order m x n and p x q respectively. Assuming n=p=2, the test data and path covered is given in table 1.

CONTROL FLOW GRAPH FOR MATRIX MULTIPLICATION :

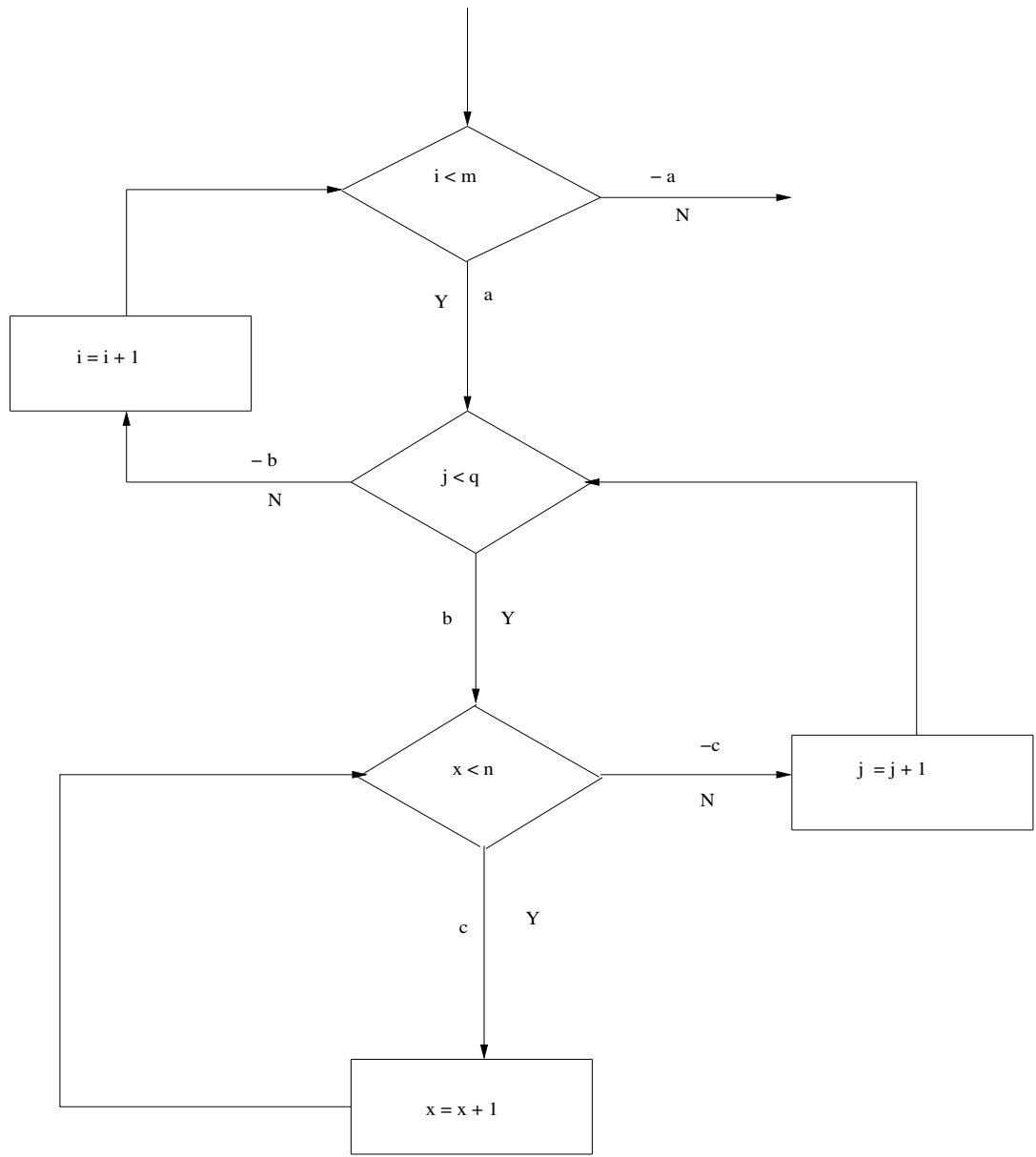

Figure 2. Control flow of matrix multiplication

## 4.2. Linear Search

Linear search takes as input an array of random integers and a random number to be searched within the array. Input: The input of the program is received from the user includes the following –

Maximum limit of the array size for a[ ]. Maximum range of domain from which elements of the array are randomly selected. Maximum number of iterations to be allowed during feasible path generation. Source code of the function Linear Search is given in Appendix I. The control flow of the graph is shown in figure 3.

Table 1. Test data generation for Matrix multiplication

| No. | m | n | p | q | a[m][n] | b[p][q] | Path Generated |
|---|---|---|---|---|---|---|---|
| 1. | 1 | 1 | 1 | 1 | [3] | [5] | a b c -c -b -a |
| 2. | 1 | 1 | 1 | 2 | [6] | [7 8] | a b c -c b c -c -b -a |
| 3. | 1 | 2 | 2 | 1 | [4 9] | $\begin{bmatrix}5\\11\end{bmatrix}$ | a b c c -c -b -a |
| 4. | 1 | 2 | 2 | 2 | [7 3] | $\begin{bmatrix}9 & 2\\1 & 6\end{bmatrix}$ | a b c c -c b c c -c -b -a |
| 5. | 2 | 1 | 1 | 1 | $\begin{bmatrix}12\\13\end{bmatrix}$ | [8] | a b c -c -b a b c -c -b -a |
| 6. | 2 | 1 | 1 | 2 | $\begin{bmatrix}5\\14\end{bmatrix}$ | [21 9] | a b c -c b c -c -b a b c -c b c -c -b -a |
| 7. | 2 | 2 | 2 | 1 | $\begin{bmatrix}9 & 2\\1 & 6\end{bmatrix}$ | $\begin{bmatrix}3\\5\end{bmatrix}$ | a b c c -c -b a b c c -c -b -a |
| 8. | 2 | 2 | 2 | 2 | $\begin{bmatrix}1 & 7\\4 & 8\end{bmatrix}$ | $\begin{bmatrix}6 & 2\\3 & 9\end{bmatrix}$ | a b c c -c b c c -c -b a b c c -c b c c -c -b -a |

CONTROL FLOW GRAPH FOR LINEAR SEARCH :

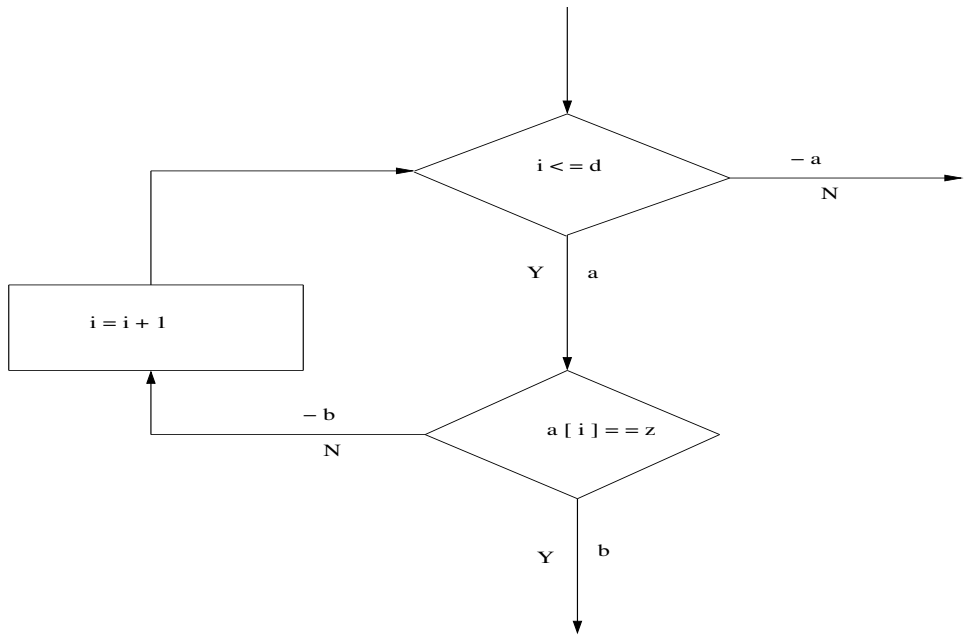

Figure 3. Control flow of linear search program

CONTROL FLOW GRAPH FOR BUBBLE SORT :

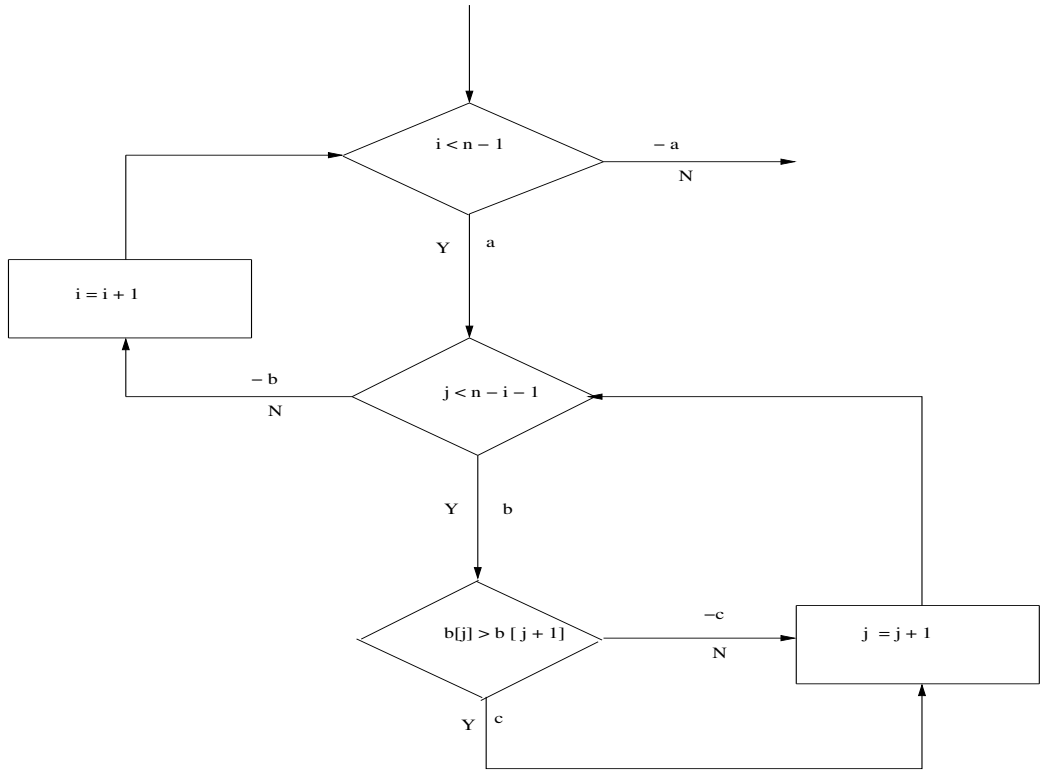

Figure 4. Control flow of bubble sort program

Test generated for Linear search: Considering an array a[l] of size l, where a[ l ] ='0,2,5,6,7' with item to be searched=7 test data and path covered is shown in table 2.

Table 2. Test data generation for Linear Search

| No. | l | a[l] | Path Generated |
|---|---|---|---|
| 1. | 1 | 0 | a -b -a |
| 2. | 2 | 0 2 | a -b a -b -a |
| 3. | 3 | 0 2 5 | a -b a -b a -b -a |
| 4. | 4 | 0 2 5 6 | a -b a -b a -b a -b -a |
| 5. | 5 | 0 2 5 6 7 | a -b a -b a -b a -b a b |

### 4.3. Bubble Sort

Source Code of the function Bubble Sort is given in Appendix. The control flow of the graph is shown in figure 4.
Input: The input of the program is received from the user which includes the following - Maximum limit of the array size for a[ ]. Maximum range of domain from which elements of the array are randomly selected. Maximum number of iterations to be allowed during feasible path generation. Test Generated for Bubble Sort: Considering an array a[ l ] of size l, where a[ l ] ='2,4,3,7,6'. Test data and path covered is shown in table 3.

International Journal of Software Engineering & Applications (IJSEA), Vol.1, No.4, October 2010

### 4.4. Implementation Algorithm

1. Generate random input and execute the instrumented program to be tested.
2. Repeat step 1 until all the feasible paths are collected in a file.
3. Filter all paths to get the unique feasible paths.
4. Repeat above steps until either the saturation level or longest path level is found and generate test data for corresponding *k*.

## 5. Experimental Results

### 5.1. Linear Search

The experimental results of linear search program given in Appendix I shows that for program with single loop the longest path value or optimal value of *k* for which no more

Table 3. Test data generation for Bubble sort

| No. | l | a[l] | Sorted a[l] | Path Generated |
|---|---|---|---|---|
| 1. | 0 | | | -a |
| 2. | 1 | 2 | 2 | -a |
| 3. | 2 | 2 4 | 2 4 | a b -c -b -a |
| 4. | 3 | 2 4 3 | 2 3 4 | a b -c b c -b a b -c -b -a |
| 5. | 4 | 2 4 3 7 | 2 3 4 7 | a b -c b c b -c -b a b -c b -c -b a b -c -b -a |
| 6. | 5 | 2 4 3 7 6 | 2 3 4 6 7 | a b -c b c b -c b c -b -a b -c b -c b -c -b a b -c b -c -b a b -c -b -a |

new feasible paths are generated is given by

$k_L$=Array Size
$k_S$= Array size + 1

We have taken 1000 domain and different array size 2,3,5,10,20,50,60,80,100. A sample data collected from our experiment for array size=3 is shown in table 4.

Table 4: Array size 3

| k | Test Cases | UFP | NFP | LLP | ETime(ms) |
|---|---|---|---|---|---|
| 0 | 100 | 1 | 1 | 1 | 1900 |
| 1 | 200 | 2 | 1 | 3 | 2100 |
| 2 | 300 | 3 | 1 | 5 | 2200 |
| 10 | 1100 | 11 | 8 | 21 | 1600 |
| 20 | 2100 | 21 | 10 | 41 | 1900 |
| 50 | 5100 | 51 | 30 | 101 | 2200 |
| 80 | 8100 | 81 | 30 | 191 | 3200 |
| 95 | 9600 | 96 | 15 | 197 | 2200 |
| 98 | 9900 | 99 | 3 | 199 | 2300 |
| 99 | 10000 | 100 | 1 | 201 | 2100 |
| 100 | 10100 | 101 | 1 | 201 | 2600 |
| 101 | 10200 | 101 | 0 | 201 | 2900 |
| 102 | 10300 | 101 | 0 | 201 | 3600 |
| 200 | 20100 | 101 | 0 | 201 | 3900 |
| 300 | 30100 | 101 | 0 | 201 | 6500 |

The graph for *k* and unique feasible path for different array size is shown in figure 5. From the graph it is clear that after array size number of iteration no more new feasible path is generated even if we increase the value of *k*. The graph for k and new feasible path is shown in figure 6.

## 5.2. Bubble Sort

The experimental results of bubble sort program given in Appendix I show that for program with two loops. The longest path value and optimal value of *k* for which no more new feasible paths are generated is given by

$$k_{Li} = k_{L(i-1)} + (arraysize-1)$$

$$k_{Li} = \sum_{i=3}^{n-1} arraysize_i + \text{base value of } k_L(arraysize=3)$$

$$k_{Si} = \sum_{i=6}^{n-1} arraysize_i + \text{base value of } k_S(6) \quad \text{where } k_{L(i-1)} \text{ is the value of } k_L \text{ in previous array size.}$$

We have taken 1000 domain and different array size 2,3,5,6,7,8,9,10,20,40. This formula for $k_L$ is applicable for array size greater than equal to 3 and that of $k_S$ is for array size greater than equal to 6. A sample data collected from our experiment for array size=20 is shown in table 5.

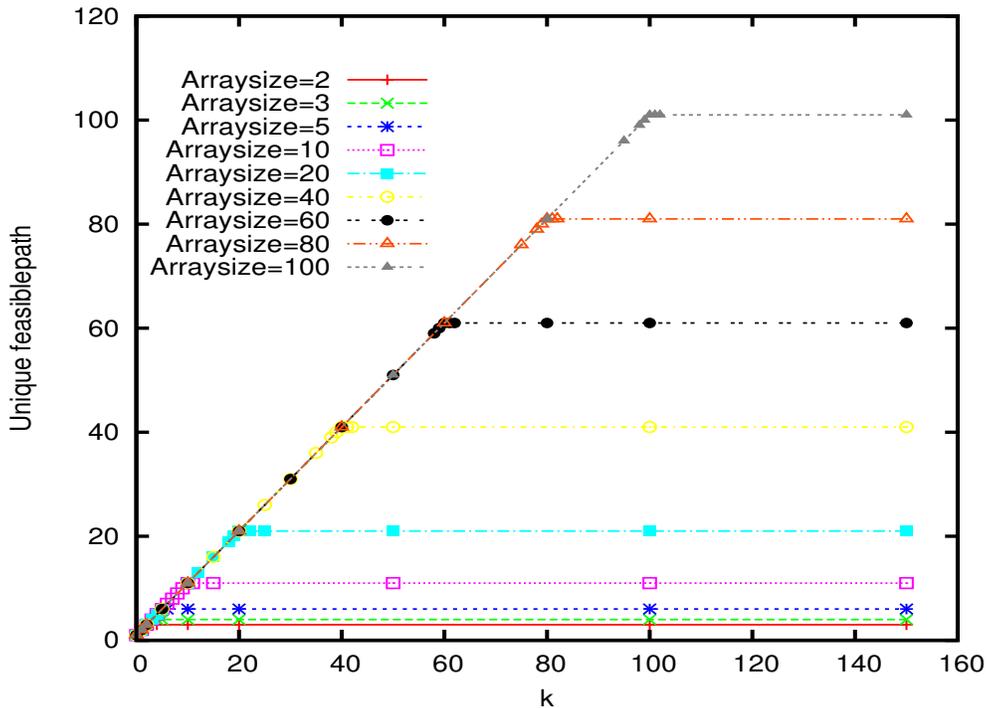

Figure 5. Linear Search: k vs unique feasible path

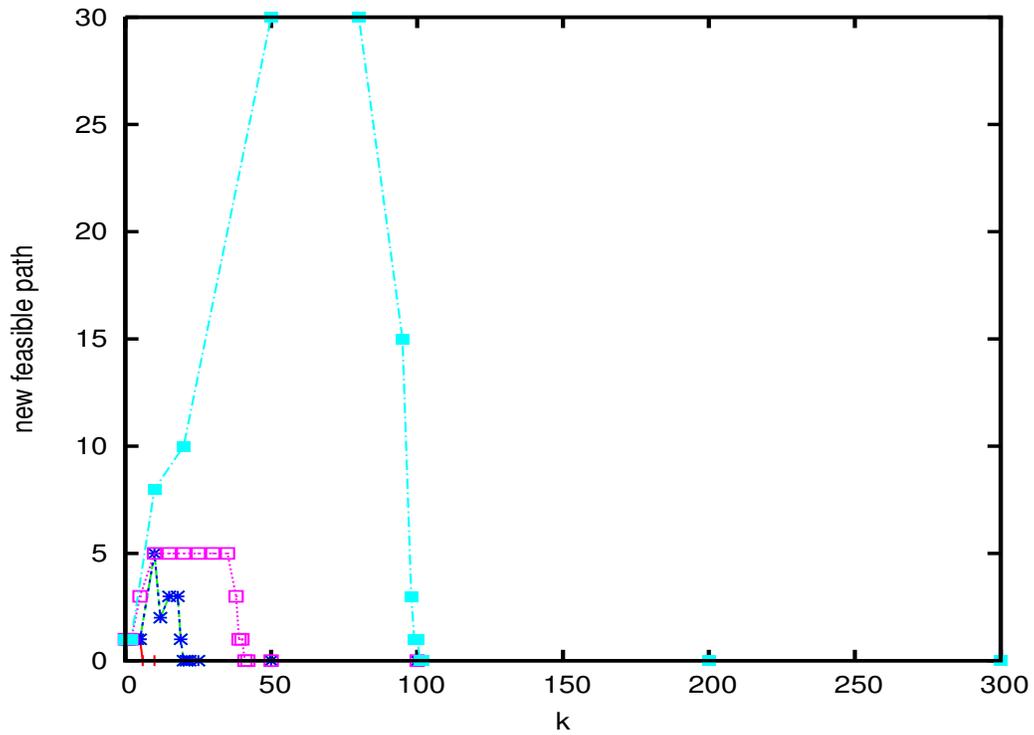

Figure 6. Linear Search: k vs new feasible path

Table 5: Arraysize 20

| k | Test Cases | UFP | NFP | LLP | ETime(ms) |
|---|---|---|---|---|---|
| 0 | 20 | 1 | 1 | 1 | 3900 |
| 1 | 40 | 2 | 1 | 5 | 1500 |
| 2 | 60 | 4 | 2 | 9 | 1000 |
| 10 | 220 | 70 | 66 | 41 | 1300 |
| 20 | 420 | 201 | 131 | 77 | 1200 |
| 30 | 620 | 320 | 119 | 97 | 900 |
| 40 | 820 | 426 | 106 | 117 | 1200 |
| 50 | 1020 | 521 | 45 | 131 | 1100 |
| 60 | 1220 | 606 | 65 | 157 | 1100 |
| 100 | 2020 | 861 | 179 | 237 | 1300 |
| 120 | 2420 | 946 | 85 | 277 | 2000 |
| 150 | 3020 | 1022 | 76 | 337 | 1600 |
| 170 | 3420 | 1045 | 23 | 377 | 1900 |
| 171 | 3440 | 1046 | 1 | 379 | 1800 |
| 172 | 3460 | 1046 | 0 | 379 | 1600 |
| 175 | 3520 | 1046 | 0 | 379 | 1300 |
| 180 | 3620 | 1046 | 0 | 379 | 2000 |
| 200 | 4020 | 1046 | 0 | 379 | 2300 |
| 220 | 4420 | 1046 | 0 | 379 | 3600 |
| 500 | 10020 | 1046 | 0 | 379 | 3500 |

The graph for *k* and unique feasible path for different array size is shown in figure 7. From the graph it is clear that after array size number of iteration no more new feasible path is generated even if we increase the value of *k*. The graph for k and new feasible path is shown in figure 8.

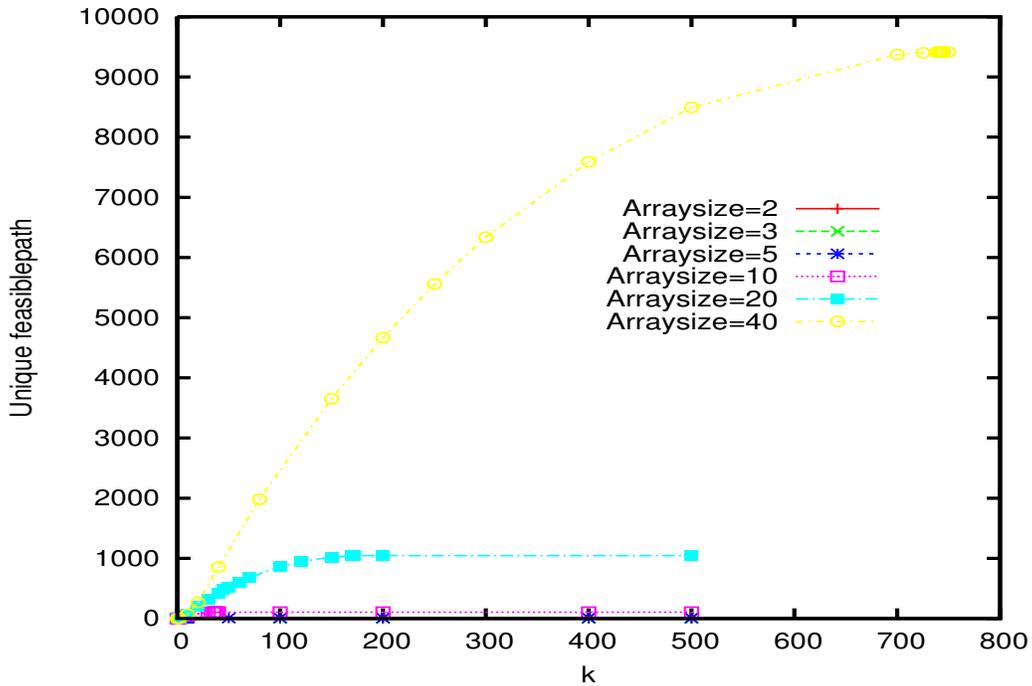

Figure 7. Bubble Sort: k vs unique feasible path

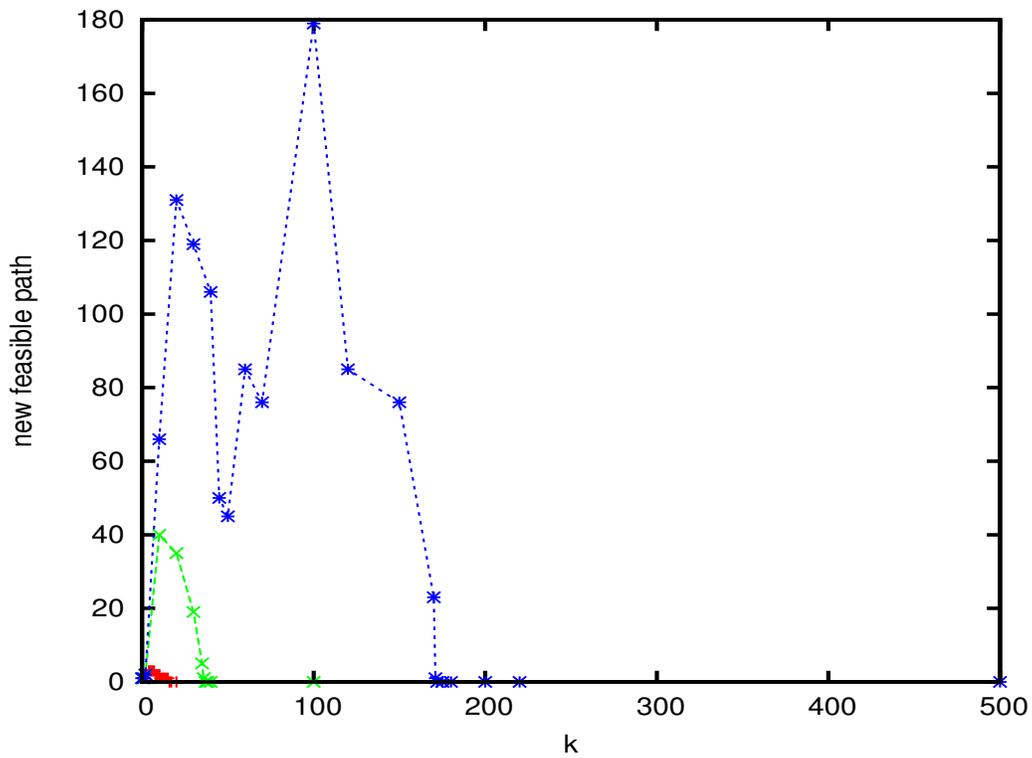

Figure 8. Bubble Sort: k vs new feasible path

## 5.3. Matrix multiplication

The experimental results of matrix multiplication program given in Appendix I show that for program with three loops. The longest path value and optimal value for which no more new feasible paths are generated is given by

$k_L$= (No of row Ist matrix * No. of Column Ist matrix * No.
 of row 2nd matrix * No. of column 2nd matrix)/No. of column Ist matrix

$k_S = k_L + 1$

We have taken 1000 domain and different matrices are
(1,2)(2,1), (1, 3)(3, 1), (1, 3)(3, 2), (1, 3) (3, 3), (2, 2)(2, 2), (3, 3)( 3, 3), (3,4)(4,6), (5, 3) ( 3, 8), (4, 5)(5, 6), (6, 3)(3, 2), (4, 4)(4, 4). The graph for *k* and unique feasible path for different matrices are shown in figure 9. From the graph it is clear that after $k_S$ number of iteration no more new feasible path is generated even if we increase the value of *k*. The graph for different size matrices and new feasible path are shown in figure 10.

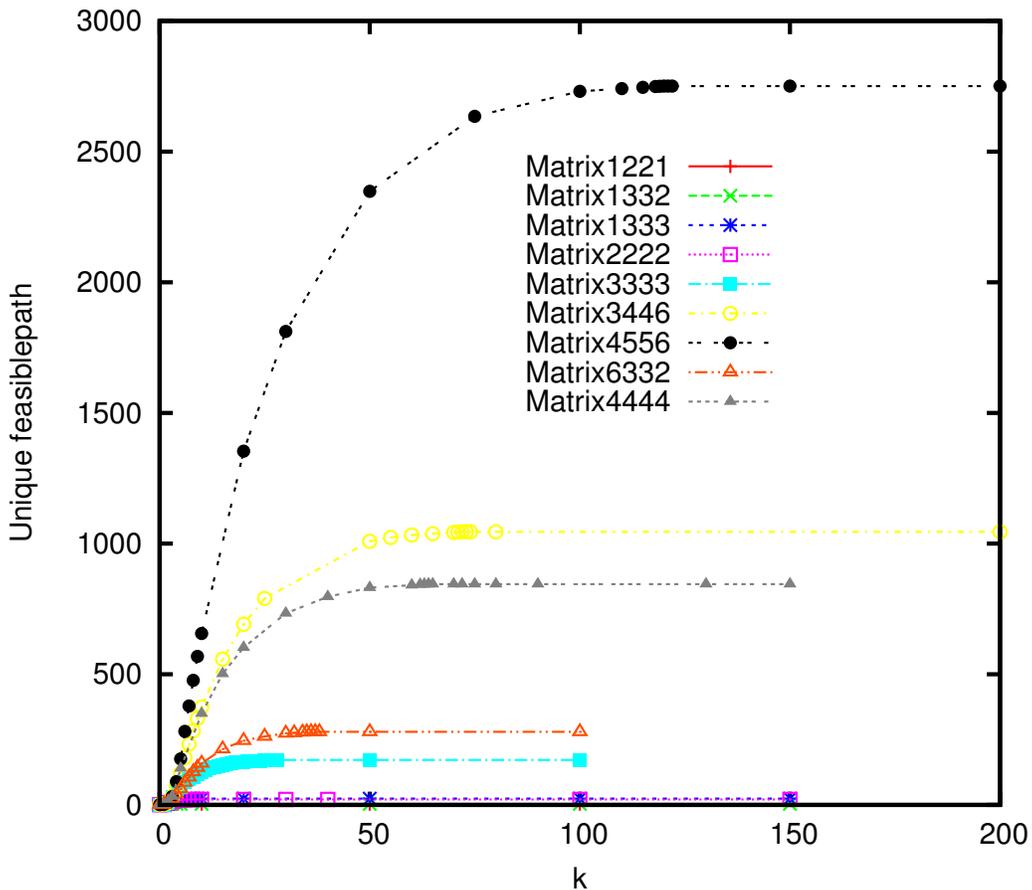

Figure 9. matrix: k vs unique feasible path

International Journal of Software Engineering & Applications (IJSEA), Vol.1, No.4, October 2010

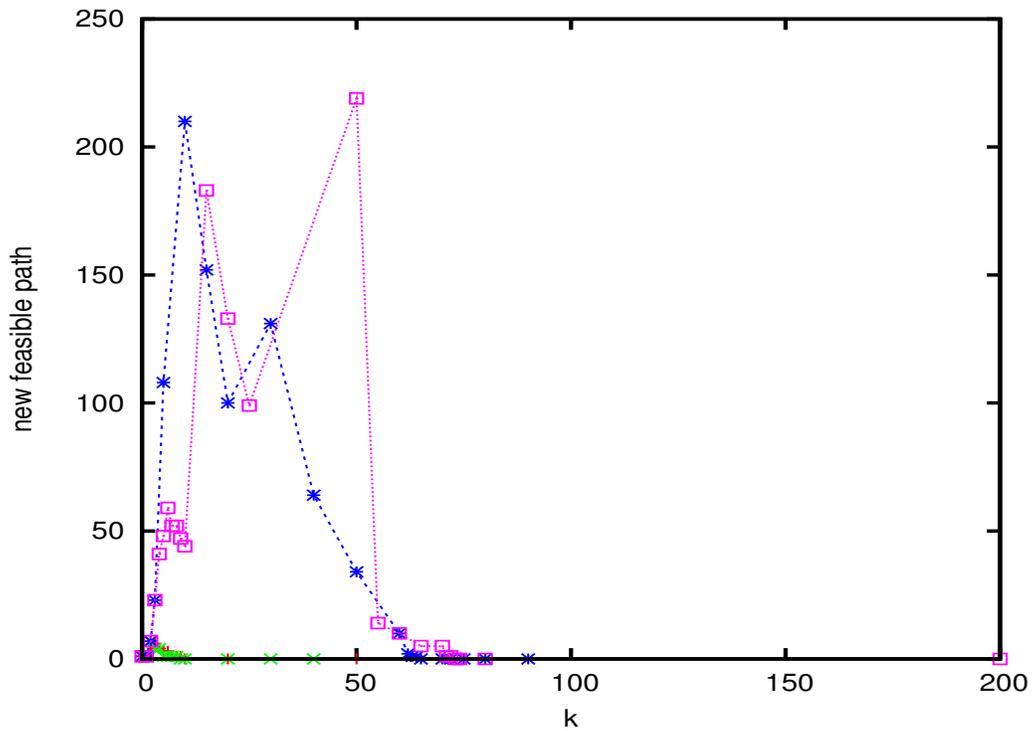

Figure 10. matrix: k vs new feasible path

A sample data collected from our experiment for matrix ( 4,4)(4, 4) is shown in table 6.

Table 6: Matrix 4444

| k | Test Cases | UFP | NFP | LLP | ETime(ms) |
|---|---|---|---|---|---|
| 0 | 64 | 1 | 1 | 1 | 1300 |
| 1 | 128 | 2 | 1 | 6 | 2000 |
| 2 | 192 | 9 | 7 | 11 | 1100 |
| 3 | 256 | 32 | 23 | 16 | 1100 |
| 5 | 320 | 140 | 108 | 24 | 1100 |
| 10 | 704 | 350 | 210 | 39 | 1500 |
| 15 | 1024 | 502 | 152 | 54 | 1200 |
| 20 | 1344 | 602 | 100 | 61 | 1500 |
| 30 | 1980 | 733 | 131 | 71 | 1200 |
| 40 | 2624 | 797 | 64 | 81 | 1500 |
| 50 | 3264 | 831 | 34 | 91 | 1500 |
| 60 | 3904 | 841 | 10 | 101 | 1200 |
| 62 | 4032 | 843 | 2 | 103 | 1600 |
| 63 | 4096 | 844 | 1 | 104 | 1600 |
| 64 | 4160 | 845 | 1 | 105 | 1400 |
| 65 | 4224 | 845 | 0 | 105 | 1500 |
| 70 | 4544 | 845 | 0 | 105 | 1600 |
| 72 | 4672 | 845 | 0 | 105 | 2300 |
| 75 | 4864 | 845 | 0 | 105 | 1600 |
| 80 | 5184 | 845 | 0 | 105 | 1500 |
| 90 | 5824+ | 845 | 0 | 105 | 2200 |

## 5.4. Merge sort

The detail experimental results of merge sort program can be seen in [12]. It is found that $k_L$ has a linear increase with array size. But $k_S$ is exponentially increased with array size. Therefore $k_L$ is taken as heuristic for this type of programming construct. The values of $k_L$ and $k_S$ for different array size of merge program is shown in table7.

Table 7: Merge sort: Array size, $k_L$,$k_S$,lmax

| Arraysize | $k_L$ | $k_S$ | Lmax |
|---|---|---|---|
| 2 | 3 | 15 | 14 |
| 3 | 8 | 49 | 20 |
| 5 | 10 | 1198 | 32 |
| 6 | 14 | 5252 | 38 |
| 7 | 15 |  | 44 |
| 8 | 16 |  | 50 |
| 9 | 18 |  | 56 |
| 10 | 19 |  | 62 |
| 20 | 41 |  | 122 |
| 50 | 100 |  | 302 |

Table 8: Linear search: Array size, $k_L$,$k_S$,lmax

| Array size | $k_L$ | $k_S$ | Lmax |
|---|---|---|---|
| 2 | 2 | 3 | 5 |
| 3 | 3 | 4 | 7 |
| 5 | 5 | 6 | 11 |
| 10 | 10 | 11 | 21 |
| 20 | 20 | 21 | 41 |
| 30 | 30 | 31 | 61 |
| 40 | 40 | 41 | 81 |
| 50 | 50 | 51 | 101 |
| 60 | 60 | 61 | 121 |
| 80 | 80 | 81 | 161 |
| 100 | 100 | 101 | 201 |

Table 9: Bubble sort: Array size, $k_L$,$k_S$,lmax

| Array size | $k_L$ | $k_S$ | Lmax |
|---|---|---|---|
| 2 | 0 | 1 | 1 |
| 3 | 1 | 2 | 5 |
| 5 | 6 | 7 | 19 |
| 6 | 10 | 11 | 29 |
| 7 | 15 | 16 | 41 |
| 8 | 21 | 22 | 55 |
| 10 | 36 | 37 | 89 |
| 20 | 171 | 172 | 379 |
| 30 | 30 | 31 | 61 |
| 40 | 741 | 742 | 1559 |

Table 10: Matrix multiplication: Matrixsize, $k_L$, $k_S$, and lmax

| Matrix size | $k_L$ | $k_S$ | Lmax |
|---|---|---|---|
| 1221 | 2 | 3 | 7 |
| 1331 | 3 | 4 | 8 |
| 1332 | 6 | 7 | 13 |
| 1333 | 9 | 10 | 18 |
| 1558 | 120 | 121 | 211 |
| 3446 | 72 | 73 | 115 |
| 4556 | 120 | 121 | 177 |
| 6332 | 36 | 37 | 73 |
| 2222 | 8 | 9 | 21 |
| 3333 | 27 | 28 | 52 |
| 4444 | 64 | 65 | 105 |

## 6. Discussion

It is fact that experimental results have several limitations to its validity. It is also not possible to set a common heuristic that will be applicable to generate test data for different programming construct. Test data generation algorithm in general is unsolvable problem.[7]. But from our experimental data we observed much similarity that can be efficient heuristic to reduce computation cost for program having variable number of loops, variable length of arrays. We experimented 3 types of programming construct( one loop and two loop with one dimensional array and three loops with two dimensional arrays. In case of linear search program, Length of the paths generated increases linearly with $k$ initially as shown in figure 5. As $k$ is increased further, at a particular value, the length of the paths become constant and does not increase any further even if value of $k$ is increased. This is the longest path criterion. It is denoted by $k_L$, value of $k$ at this point. In the graph for array size 100, the length of the longest path become constant at k=100 and therefore $k_L$ =100. It is observed that for a particular value of $k$, a saturation point is achieved after which no more new possible paths are generated. At $k$=101, for array size 100 we attained a saturation with 101 feasible paths, which means that even if we increment the value of $k$, the number of feasible paths remains constant. Let $k_S$ denote the value of $k$ at this point. In case of bubble sort program, length of path generated increases quadratically as the value of array size increases as shown in figure 7. The value of longest path length becomes constant at $k_L$ number of iteration and the number of paths generated becomes onstant at $kS$ number of iterations. For example for array size=20 as shown in table 5, the value of the longest path become constant at $k$ = 170 and therefore $k_L$ =170. No new paths are generated after $k$ = 171, so the value of $k_S$=171. Similarly we got $k_L$ and $k_S$ for matrix multiplication program as shown in the table 6. Here $k_L$=63, and $k_S$=64. It can be observed from table 7 that the value of $k_L$ increases linearly with increase of array size. Due to exponential increase of $k_S$ we are taking only $k_L$ for large array size to save the computational time. Therefore test data generation tool we may either use $k_L$ number of iteration or $k_S$ number of iteration depending on type of programming constructs. $k_S$ ensure all path coverage and $k_L$ ensures all path coverage containing the longest path. We may satisfy the coverage with $k_L$ iteration if the value of $k_S$ exponentially increases with the increase of the array size. The values of $k_L$ and $k_S$ for different array size for programming construct linear search, bubble sort, and matrix multiplication are shown in tables 8, 9, 10 respectively. It has been observed that value of $k_L$ can be computed for any array size and programming construct with less effort as compared to $k_S$. One observation from our plotted data is that there is always saturation on number of path that can be generated in a loop construct. Taking less number of iteration we can not get better coverage. Taking more iteration is costly. Therefore our experimental data gives us the lower bound on number of iteration to be allowed for either longest path coverage or all path coverage and that can be used as heuristic for that type of programming construct. We have found from our practical observation that the execution time of the program for greater array sizes increases manifolds. Also it depends on the configuration of the machine. We used Xeon (IBM System X3650) with 2GB RAM and Red Hat Enterprise Linux 5.0 operating system for our experiments.

## 7. Conclusion and Future Work

In this paper, we have described techniques for finding minimum number of iterations to be adopted to get test data for all path coverage or longest path coverage for programs with loops and arrays of variable length. Our sample represents the basic programming constructs of almost all sequential programs. By studying the behavior of our prototypes on those for arrays of different dimensions and sizes with variable user defined number of iterations(k), we found that

1. For a particular value of $k$, a saturation point is achieved after which no new unique feasible paths are generated. We term this value of $k$ as $k_S$.
2. For a particular value of $k$, a maximum path length is obtained for each array size. This length remains constant thereafter even if $k$ is increased further. This value is termed as $k_L$.
3. It is observed that for almost all type of programming construct the value of $k_L$ increase linearly with the increase of array size.

$k_L$ may be applicable for those programs where number of paths increases exponentially when we increase the array size. In that situation, it is not feasible to determine $k_S$. $k_L$ gives us longest path coverage , $k_S$ gives us all path coverage. We have chalked out relations which predict the value of $k_S$ and the length of the longest path(lmax) for a given array size. But for merge program we restrict our heuristic to $k_L$ as the value of $k_S$ increases exponentially with increasing array size. The relations for $k_S$ satisfy the rigorous all paths criterion. The relationship found between array size, $k_L$, and $k_S$ are independent of the domain. Given an array size for a program We can determine the value of $k_S$, and $k_L$ of that program. It has been observed that value of $k_L$ can be computed for any array size and programming construct with less effort as compared to $k_S$. The various possibilities of inputs are taken as test input to observe the abilities to improve fault detection by those test input. Our results suggest that more experiments can be done for different types of commonly found programming constructs. The minimum number of iteration required for all path coverage or longest path coverage can be listed as a heuristic table for test data generation problem of programs having loops and arrays. Our method is ignoring all infeasible paths and no constraint solving is required. The all paths are filtered to get unique paths using shell script. Our model is less costly because it avoids constraint solving and no time spent on infeasible path detection. We have found from our practical observation that number of paths increases with array size for some program linearly, for some program quadratically, and for some program exponentially. Accordingly we will take heuristic either $k_L$ or $k_S$. The behavior of $k_S$ should be observed with more examples to obtain a greater precision. For that we require to experiment many different types of sample programs. In future, a generalized heuristic table can be formed for different types of programming constructs with real life examples for $k_L$ and $k_S$. Our method is seems to be good provided we can predict the minimum number of iterations required to find the all feasible paths from a heuristic table. Our testing method is useful for unit testing.

### ACKNOWLEDGEMENTS

The authors would like to thank the anonymous reviewers for their helpful comments and suggestions.

Annexure I

**Source Code of the function Linear search:**

```
int linear_search(int a[],int d,int z){
        int i,d,z,f;
        for(i=1;i<=d;i++){
        if(a[i]==z)
        f=1;

    else
      f=0;
    }
   return f;    }
```

**Source Code of the function Bubble Sort:**

```
   void Bubble_sort(int b[],int n){
       int i,j,temp;
       for(i=0;i<n-1;i++){
          for(j=0;j<n-i-1;j++){
             if(b[j]>b[j+1]){
                                    temp=b[j];
                            b[j]=b[j+1];
                                b[j+1]=temp;
                          } } } }
```

**Source Code of the function Matrix multiplication:**

```
void Matrix_mult(int a[][],int b[][],int m, int n, int p,int q)
{       int c[20][20],i,j,x;
         for(i=0;i<m;i++){
         for(j=0;j<q;j++){
                  c[i][j]=0;
                for(x=0;x<n;x++){
                c[i][j]=c[i][j]+ a[i][p]*b[p][j];
                } } } }
```

**Authors**

**H. Tahbildar** Received his B. E. degree in Computer Science and Engineering from Jorhat Engineering College, Dibrugarh University in 1993 and M. Tech degree in Computer and Information Techno logy from Indian Institute of Technology, Kharagpur in 2000. Presently he is doing Ph.D and his current research interest is Automated Software Test data generation, Program Analysis. He is working as HOD, Computer Engineering Department, Assam Engineering Institute, Guwahati, INDIA.

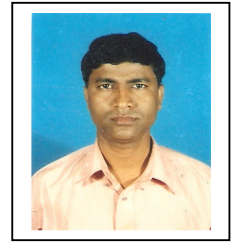

**B. Kalita**: Ph.D degree awarded in 2003 in Graph Theory. At present holding the post of Associate Professor, Department of Computer Application, Twenty research papers have got published in national and international level related with graph theory, Application of graph theory in VLSI design, software testing and theoretical computer science. Field of interest: Graph theory, VLSI Design, Automata theory, network theory, test data generation etc. Associated with the professional bodies, such as Life member of Indian Science Congress association, Life member of Assam Science Society, Life member of Assam Academy of Mathematics, Life member of Shrimanta Sankar deva sangha ( a cultural and religious society). Delivered lecture and invited lectures fourteen times in national and international level.

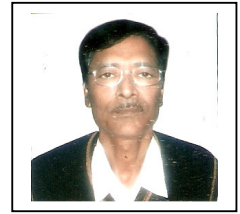